# Colossal Dielectric Permittivity and Superparaelectricity in phenyl pyrimidine based liquid crystals

Yuri P. Panarin[1,2#], Wanhe Jiang[3#], Neelam Yadav[1#], Mudit Sahai[1,4], Yumin Tang[5], Xiangbing Zeng[5], O. E. Panarina[1], Georg H. Mehl[3*], Jagdish K. Vij[1*]

[1]Department of Electronic and Electrical Engineering, Trinity College Dublin, The University of Dublin, Dublin 2, Ireland
[2]Department of Electrical and Electronic Engineering, TU Dublin, Dublin 7, Ireland
[3]Department of Chemistry, University of Hull, Hull HU6 7RX, UK
[4]Department of Physics, Birla Institute of Technology and Science, Pilani, India
[5]Department of Materials Science and Engineering, University of Sheffield, Sheffield, S1 3JD, UK

A set of polar rod-shaped liquid crystalline molecules with large dipole moments (μ > 10.4-14.8 D), their molecular structures based on the ferroelectric nematic prototype DIO, are designed, synthesized, and investigated. When the penultimate fluoro-phenyl ring is replaced by phenylpyrimidine moiety, the molecular dipole moment increases from 9.4 D for DIO to 10.4 D for the new molecule and when the terminal fluoro- group is additionally replaced by the nitrile group, the dipole moment rises to 14.8 D. Such a replacement enhances not only the net dipole moment of the molecule, but it also reduces the steric hindrance to rotations of the moieties within the molecule. The superparaelectric nematic (N) and smectic A (SmA) phases of these compounds are found to exhibit colossal dielectric permittivity, obtained both from dielectric spectroscopy, and capacitance measurements using a simple capacitor divider circuit. The electric polarization is measured vs. the field ($E$). However, no hysteresis in $P$ vs. $E$ is found in the nematic and smectic A phases. The colossal dielectric permittivity persists over the entire fluidic range. The experimental results lead us to conclude that these materials belong to the class of superparaelectrics (SPE) rather than to ferroelectrics due to the absence hysteresis and linear P vs E dependence. The synthesized organic materials are the first fluids for which superparaelectricity is discovered and furthermore these show great potential for the applications in supercapacitors used in storing energy.

# 1. Introduction

In electronics industries, the need for increasing the device density goes hand in hand with miniaturization of electronic structures that continue working efficiently with decreasing size to the nanoscale level. The micro and nano electronic devices thus require replacement of the conventional dielectric materials with those exhibiting colossal dielectric permittivity (CP). CP has been observed in solid state ferroelectric and superparaelectric (SPE) pervoskite related materials. The concept of superparaelectricity, introduced recently [1] is somewhat analogous to superparamagnetism (SPM). The electric polarization induced by the field is identified as superparaelectric with the basic property of the materials having CP. In an analogous case of magnetism, the theory of the ordered states based on isolated spins has been worked out. A superparamagnetic state includes small clusters of ordered spins with adequate magneto-crystalline anisotropic energy that maintain stability against thermal fluctuations. However no theory is yet established for the solid SPE though polar clusters of varying sizes may exist to form a polar order.

These range mainly from the transition metal oxides [2,3] to two-dimensional nanosheet hybrids, the latter are based on reduced graphene oxide. [4] These systems are found having varying dependencies of the permittivity on temperature and frequency. For example, some SPE materials show CP independent of frequency up to the GHz range [5], while others such as the relaxor ferroelectrics show strong resonance-type frequency dependence of the permittivity [1]. In the paraelectric state, [1, 5] the solid state materials show softening of the permittivity with temperature increasing (soft mode), and in some cases the permittivity rises with increase in temperature. Due to these unusual properties for the new phenomenon, the paraelectric state is described as superparaelectric (SPE).

Although there are no strict definition of SPE state which might have different physical origin however all SPE materials have two common features (i) Colossal Permittivity CP [6] and (ii) paraelectric type (i.e. hysteresis-free linear) P-E response.

The first examples of fluids that exhibit colossal dielectric permittivity are the recently observed ferroelectric nematics,[7, 8, 9,10] characterized by (i) extremely large dipole moments ($\mu$ ~10 D), (ii) colossal dielectric permittivity ($\varepsilon'$ ~ 10,000) and (iii) the high spontaneous polarization ($P_s$ ~ 5 µC/cm$^2$) [11,12]. Such high dielectric permittivity explained by two different theoretical models [13,14] and [15,16,17]. Interestingly, some materials show optical activity from mirror symmetry breaking [18,19,20] in both the ferroelectric and paraelectric nematic phases. Though several new organic ferroelectric compounds [21,22,23,24,25,26,27,28,29] continue being reported and discussed in the literature every week, many critical issues need addressing prior to fabricating and launching devices for applications. Though the structure property relationship of ferroelectric nematics has not yet been worked out, two major reasons for the emergence of this phase have been found. These are (i) the magnitude of the molecular dipole moment and (ii) the relative spatial distribution of dipole moments of groups and or the charge distribution created within the molecule. Finally, the short and the long-range intermolecular interactions of the dipole moments of molecules lead to the mesoscale synergistic properties that need investigations.

In the quest for finding a novel class of high-dielectric constant organic materials based on the prototype molecule DIO [7], Table 1, the molecular design includes enhancing its dipole moment. Large molecular dipole moment is also the pre-requisite for obtaining large dielectric permittivity. In addition, the short-range Kirkwood Frohlich correlation parameter of the dipole moments [30] also plays an important role. In this paper, new molecules based on the substituted phenylpyrimidine motif are designed, synthesized, and investigated. Polarizing optical microscopy, X-ray scattering, electrical studies, and dielectric spectroscopy are used to characterize different phases of two new compounds. Large values of the dielectric permittivity imply dipolar orientation of the polar clusters at a microscopic level in the medium. A proof for the existence of the super paraelectricity in these materials is presented in terms the absence of saturated polarization.

## 2. Experimental Section

### 2.1. Sample preparation for dielectric, optical and electro-optical studies.

For achieving planar alignment where required, Indium tin oxide (ITO) coated glass substrates are spin coated with RN 1175 (Nissan chemicals, Japan) and polymerized at a temperature of 250 $^{o}$C for 1 hour. The coated surfaces are subsequently rubbed with a rotating commercial rubbing machine. While the homeotropic cells are coated with AL60702 (JSR Korea) and polymerized at 80 $^{o}$C for 15 minutes and at 110 $^{o}$C for 15 minutes respectively. Commercial cells procured from E.H.C ltd., Japan are also used for some of the measurements. The cell thickness was controlled by Mylar spacers of different thicknesses and it was measured by optical interference technique. LC cells of these samples are studied using polarizing optical microscope (Olympus BX 52) equipped with an INSTEC's hot stage. The temperature is controlled by Eurotherm 2604 and system designed to obtain temperature stabilization within ±0.02$^{o}$C

### 2.2. Differential Scanning Calorimetry

The DSC is investigated using Perkin Elmer Differential Scanning Calorimeter DSC 4000, using aluminum pans and calibrated against the indium standard. DSC results are cited as the onset temperatures of the second heating and cooling curve. The heating rate, if not cited is 10 °C/min.

## 2.3. Birefringence Measurements

The birefringence measurements are made with an optical spectral technique [31] using a planar homogenous aligned 25 µm thick cell. The transmittance (T) spectra of the cell from achromatic light source is measured using Avantes AvaSpec-2048 fiber spectrometer as a function of temperature. The transmittance T of a homogeneous planar aligned cell is given by:

$$T = A \sin^2 \left( \frac{\pi \cdot \Delta n(\lambda) \cdot d}{\lambda} \right) + B \qquad (9)$$

where $A$ is the amplitude factor, $B$ is leakage offset, $d$ is the cell thickness and $\Delta n(\lambda) = k \cdot \frac{\lambda^2 \cdot \lambda^{*2}}{\lambda^2 - \lambda^{*2}}$ is the birefringence dispersion governed by the extended Cauchy equation. The birefringence data for a wavelength of 550 nm are calculated using the software developed in our laboratory and the results so obtained are plotted in Fig.2.

## 2.4. X-ray diffraction

Simultaneous small and wide-angle X-ray Scattering (SAXS/WAXS) experiments were carried out at station I22 of the Diamond Light Source. The samples are inserted in 1mm diameter glass capillary and is placed in a magnetic cell of field strength of ~0.5 Tesla in the horizontal direction placed on the top of a Linkam heating stage for temperature control. The X-ray wavelength used was 1.0 Å and 2D SAXS/WAXS patterns respectively were collected with two Pilatus detectors.

## 2.5. Dielectric Spectroscopy

Dielectric relaxation measurements over a frequency range 1 Hz–10 MHz were made using a broadband Alpha High Resolution Dielectric Analyzer (Novocontrol GmbH, Germany). The glass substrates coated with a low sheet resistance (5 Ω/□) ITO electrodes were used to make cells. The low sheet resistance is used such that the peak frequency arising from the sheet resistance of ITO in series with the capacitance of the cell is shifted to a frequency much higher than 1 MHz. A prior measurement of the capacitance of the empty cell is made. The measurement is carried out under the application of weak voltage 0.1V applied across the cell. Temperature of the sample is stabilized to within ±0.05 °C. The dielectric spectra are analyzed using the Novocontrol WINDETA program.

## 2.6. Electrical Measurements

An electric signal from an Agilent 33120A signal generator amplified by a high voltage amplifier (TReK PZD700) is applied across the cell. The output signal from resistive (1 kΩ) or capacitive (2 μF) loads was monitored by a digital oscilloscope.

## 3. Results and Discussion

## 3.1. Molecular design of materials

This paper reports a study of the two newly synthesized phenylpyrimidines: WJ-16 and WJ-18 (Table 1). Here a fluorophenyl group of DIO is replaced by a narrower pyrimidine group and this replacement increases the molecular dipole moments from 9.4 D for DIO to 10.4 D for WJ-16.

Table. 1. The chemical structures of molecules WJ-16, WJ-18 and of the prototype DIO with the phase transition temperatures determined using polarizing microscopy. The dipole moments are calculated using DFT B3LYP/ 3-21G with the Gaussian View 9.0. Here N and SmA denote superparaelectric 9sp0 nematic and smectic A phases.

| Name of Compound | Molecular structure | Phases and Transition Temperature (°C) | Dipole Moment (D) |
|---|---|---|---|
| **DIO** | 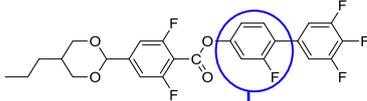 | $N_F$ 66.8 $SmZ_A$ 83.5 N 173.8 I | 9.4-9.5[7,32] |
| **WJ-16** | 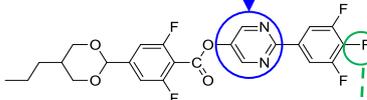 | Cr 79.3 SmA 110.5 N 198.6 I | 10.4 |
| **WJ-18** | 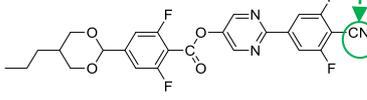 | Cr 125.6 N 230 Decomposes | 14.8 |

In WJ-18, the terminal fluoro-group is additionally replaced by the cyano-group. This increases its molecular dipole moment from 9.4 D for DIO to 14.8 D. It should be mentioned that this is not highest dipole moment as compere to giant 33.5 recently reported [23]. Details of the design, synthesis and chemical characterization of WJ-16 and WJ-18 and their crucial intermediates are given in the supplemenF[21tary Information (ESI, S1- S36; S41 gives DFT calculated structures). According to the $^1$H-NMR spectroscopic data, WJ-16 does not contain isomers associated with a substitution pattern of the dioxane group, recently discussed for DIO. For WJ-18 the amount of isomers is at the detection limit of ~1-2% (ESI, S13, S30). [33, 34] The dipole moments of the molecules are calculated using DFT Gaussian software suite at the B3LYP/ 3-21G (D, P) level. The use of pyrimidine group instead of the fluorophenyl motif reduces the steric hindrance to the rotations of groups and it enables increased π-conjugation of the aromatic fragments, resulting in an increased planarity of the terminal aromatic rings. On cooling from the nematic phase WJ-16 in addition exhibits SmA phase. The transition temperatures are given in Table 1. Above the crystalline phase in WJ-18, nematic phase is observed, whereas the compound decomposes prior to reaching its N-Iso transition temperature. Differential scanning calorimetry (DSC) scanned at 10 ˚C min$^{-1}$ on heating shows WJ-16 melting at 119.5 ˚C into the LC state, and it then transforming to the isotropic phase at 200.1 ˚C, with a transition enthalpy (ΔH) of 0.66 Jg$^{-1}$ (0.21 kJmol$^{-1}$) (Figure S37, ESI). On cooling, the liquid crystalline phase emerges at 199.0 ˚C (ΔH: -0.67 J g$^{-1}$; -0.33 kJ mol$^{-1}$), thermodynamically unstable (monotropic) additional LC phase emerges at 111.1 ˚C (ΔH: -0.22 J g$^{-1}$; -0.21 kJ/mol), prior to its crystallisation. WJ-18 shows the phase transition from the N to the crystalline state at 119.8 ˚C (Figure S38, ESI). Since the dipole moments of molecules are increased significantly as compared to DIO, the dielectric permittivity for both WJ-16 and WJ-18 becomes colossal as discussed below.

**3.2. The Optical Textures of Liquid crystalline phases.**

For investigating the optical textures, we use commercial (E.H.C. Co. Ltd. Japan) and laboratory fabricated "uncoated" ITO electrodes cells of thicknesses varying from 2 to 25 μm, Both planar and homeotropic alignment is achieved depending on the alignment layer put on to the cell electrodes. The fabrication of cells in the laboratory involves spin-coating surfactant on ITO surfaces of substrates. The handmade uncoated (i.e. without alignment layers, bare electrodes) cells are also used. Fig. 1 shows textures obtained from polarizing optical microscopy (POM) of 9 μm cell filled with WJ-16 under different alignment configurations.

The textures of planar aligned cells where the rubbing direction (R) makes an angle of α = 45º with the polarizer/analyzer (P/A) axis are identical both in the nematic and the smectic phases except for a change in the color (Figure S39, ESI). This arises mainly from a large increase in the birefringence that occurs at the N-SmA transition temperature, under cooling. Some differences in the optical textures in the N and SmA phases are observed as the cell is rotated by a small angle, e.g. α ≈ 5°, the rotation of the axis is indicated by a blue arrow drawn in Figure 1B. To explore it further, the POM textures of a planar aligned cell were recorded with the rubbing direction nearly parallel to the polarizer axis. Here, the observed texture in the N phase (Fig. 1A) highly fluctuates and is without domains – a typical feature of conventional nematics. While texture in the SmA phase is non-fluctuating and has well defined domains (Figs. 1 B, C), two typical features of smectics. It is worth noting that the textures of parallel rubbed cells do not show 'twisted domains' - characteristically found in ferroelectric [35,36] nematic phase ($N_F$) of DIO, but the texture appears domain-less rather like that of the ordinary nematics.

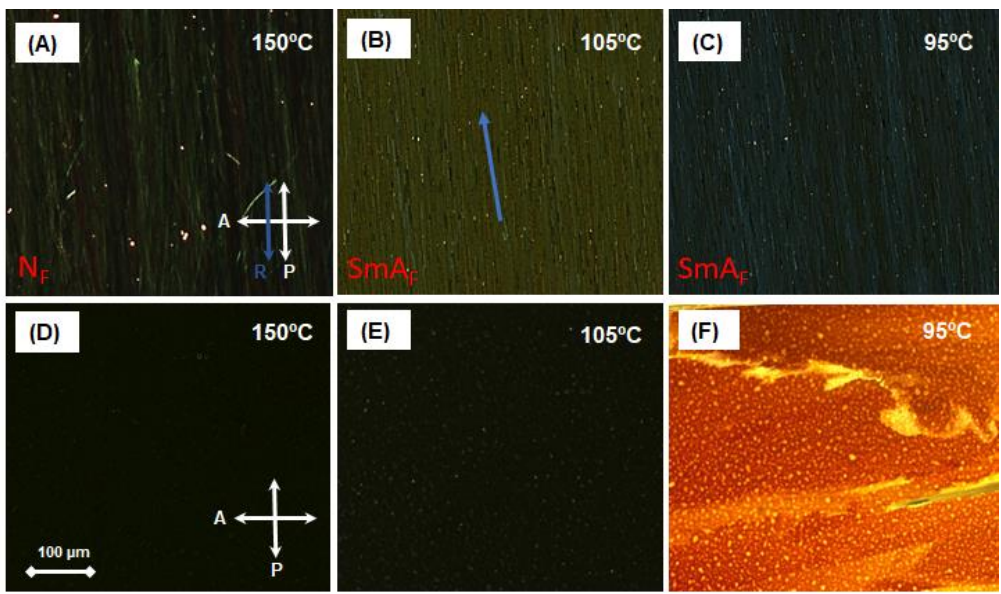

**Figure 1.** POM textures of a homogeneously aligned planar cell **(A-C)** filled with **WJ-16** with the rubbing direction (in blue) lying parallel to the polarizer, and bare ITO cell using the same compound **(D-F)** at temperatures of 150 ˚C, 105 ˚C, and 95 ˚C. Here A and P refer to the analyzer and the polarizer, and R in blue refers to the rubbing direction in a planar aligned cell. The scale bar shown in 2D is of the length 100 µm.

Both WJ-16 and WJ-18 filled in commercial homeotropic cells show perfect homeotropic texture, in contrast to the texture of the $N_F$ phase of DIO [7,19‑11] where the partial *Schlieren* texture is observed in a homeotropic aligned cell. WJ-16 when filled in uncoated cells ( i.e. bare ITO electrodes) shows perfect homeotropic texture with significantly large extinction in the nematic and SmA phases (Figure 1D, E). On further cooling the sample cell to a temperature of 95 ˚C, the texture transforms to a non-homogeneous planar one (Figure 1F). This surprising property of obtaining homeotropic texture with bare electrodes allows us to record dielectric spectra in homeotropic configuration without the alignment layers. This is fortuitous property since the alignment layers may introduce uncertainty in the measurements of dielectric permittivity. WJ-18 displays only the nematic phase. However, in uncoated cells, WJ-18 unlike WJ-16 does not show homeotropic alignment but gives a non-homogenous planar texture (Fig. S40, ESI). We note that the isotropic phase of WJ-18 is unreachable due to its anticipated thermal decomposition occurring at temperature of 230 °C.

To characterize the LC phases, birefringence of WJ-16 is measured in a planar-aligned cell of 25 μm cell thickness and calculated for a wavelength of 550 nm. Fig. 2 shows that the birefringence increases with decreasing temperature, and a significant step increase in birefringence occurring at the N to SmA transition temperature is observed. The behavior of these phases relates to the observed increase in the nematic order parameter $S$. This is verified by fitting the birefringence data to the empirical Haller equation [37] (shown as red line in Fig. 2) with $S_{\Delta T} = S_0(\Delta T)^\gamma$, where $\Delta T = T_{N\text{-Iso}} - T$, $\gamma = 0.25$. On using the wide-angle X rays scattering (WAXS) set-up and the sample alignment achieved by external magnetic field, we obtain $\gamma = 0.28$ [Figure S42, Supplementary Information]. At the N to SmA phase transition temperature, the orientational order parameter increases significantly by as much as 0.16. This is a much larger step increase in the order parameter compared to the calamitic and bent-core LC systems reported in the literature. A comparison of the plots of the birefringence vs. temperature of WJ-16 with DIO shows that only a single nematic phase exists in WJ-16 unlike DIO, where more than one nematic phase is observed.

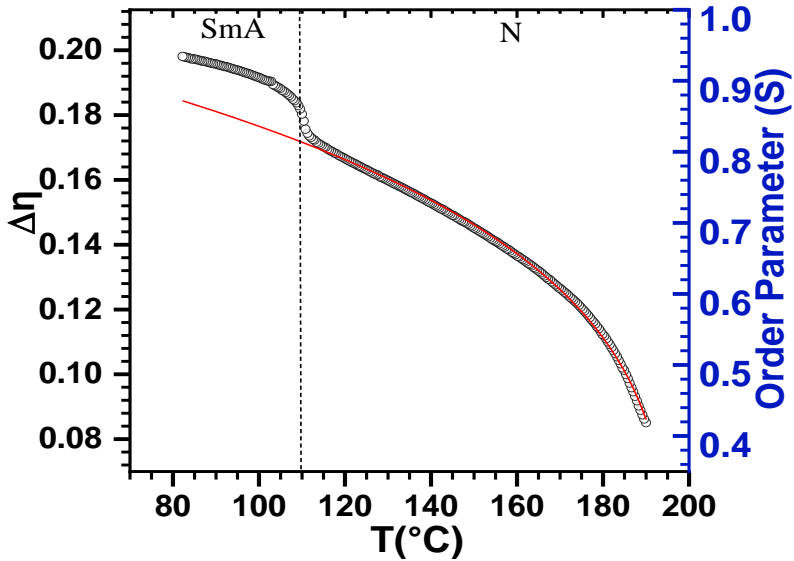

**Figure 2**. The birefringence ($\Delta n$) plot of **WJ-16** (with scale on the left Y-axis) and the order parameter (scale on the right Y-axis) as a function of temperature for a wavelength of $\lambda = 550$ nm obtained using a homogeneous planar aligned 25 μm thick cell. The red thin line is a fit of the Haller's equation to the birefringence data. Interestingly a significant increase in the order parameter (~0.16) occurs at the N to SmA transition temperature.

### 3.3. X-ray diffraction Study

To characterize the nanoscopic assembly behavior of WJ-16, small (SAXS) and wide angle (WAXS) X-ray scattering studies are carried out on magnetically aligned samples using synchrotron radiation. The nematic phase observed at temperatures between ~200 ˚C and 111 ˚C is characterized by a broad, but well oriented scattering peaks observed in both the small and the wide-angle regions (Figure 3A). The SAXS peak maxima lies in the horizontal direction, i.e., along the direction of the magnetic field. These observations indicate the average distance between the centers of molecules is 23.5 Å at 130 ˚C which is close to the molecular length found from molecular modelling (24 Å) and is consistent with the expectation that the molecular axes are oriented along the field direction. The main WAXS peak is observed in the vertical direction, with a corresponding *d*-spacing of ~4.5 Å. This again is consistent with the expected average distance between the molecules in the lateral direction. On cooling from the nematic phase, a sharp Bragg diffraction peak is observed at a temperature of ~111 ºC, arising from a formation of the smectic phase. On further cooling, the sample goes to the crystalline state where multiple sharp Bragg peaks are seen in both SAXS/WAXS regions. The SAXS peak of the smectic phase has a corresponding *d*-spacing of 23.9 – 24.1 Å ($q \sim 0.262$ Å$^{-1}$). This again agrees with the molecular length, and it confirms that the phase is of the SmA type, i.e., the average direction of the molecules is parallel to the smectic layer normal.

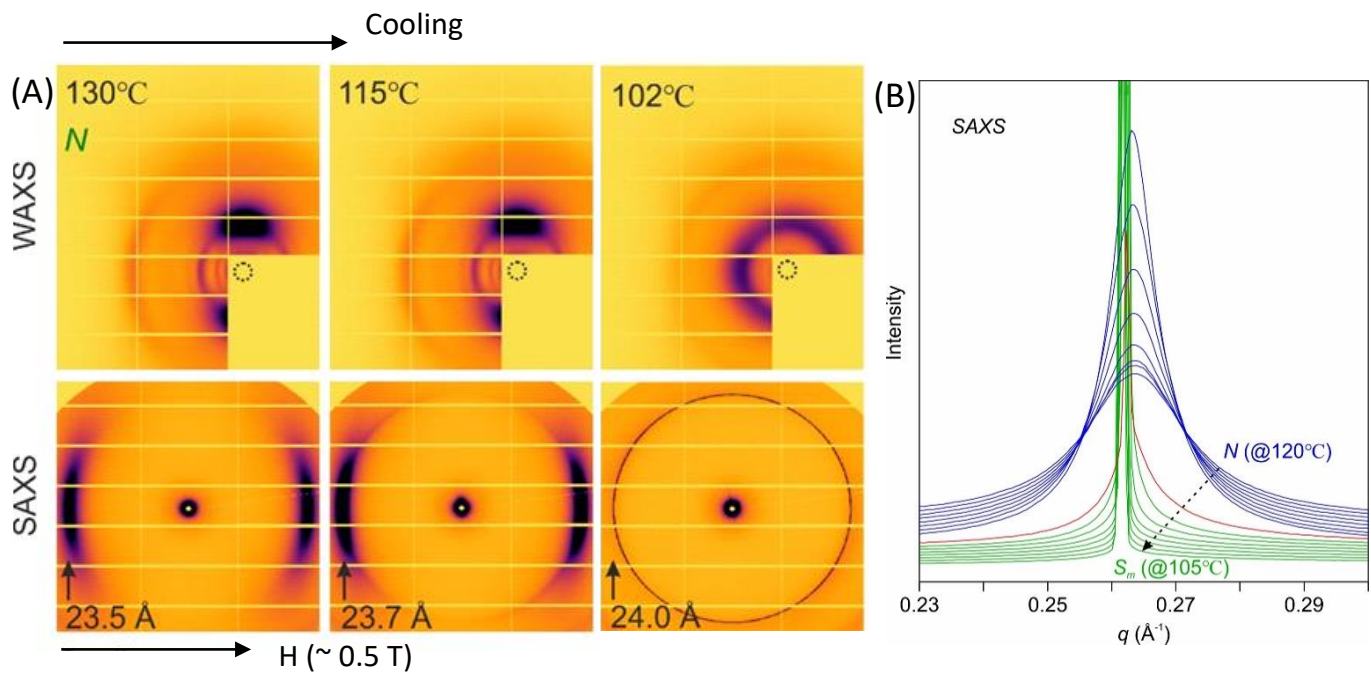

**Figure 3.** **(A)** Simultaneous SAXS/WAXS patterns of phases formed by **WJ-16** on cooling from 140 ºC to 85 ºC at a rate of 2 ˚C /min, with *d*-spacings of the scattering/diffraction peaks marked in SAXS. The WAXS beam center is indicated by a dashed circle, 3(**B**) 1D plot of the SAXS intensities as a function of *q* for different temperatures (120 -105 °C), N is indicated in blue, SmA in green, the phase transition temperature N -SmA is shown in red.

The distinct sharpening of the small angle scattering peak with decreasing temperature at the transition from the N to SmA is shown in Figure 3B. The *d*-spacings observed in the N phase of WJ-16 are like those observed in DIO [7]. On the basis of the FWHMs of the SAXS and the WAXS peaks, we calculate the correlation lengths in the nematic phase both in longitudinal and lateral directions. The lateral correlation length is essentially constant in the entire N phase (~3.2 Å), thus confirming the liquid-like short range positional order of N-phase. However, the longitudinal correlation length decreases, first rapidly just above the SmA phase, from ~210 Å at 116 °C to ~100 Å at 122 °C, but then slowly to -50 Å at 150 °C and ~23 Å (single molecular length) just prior to the occurrence of the N-Iso transition (Figures. S43 and S44, ESI). The observed simultaneous SAXS/WAXS patterns of WJ-18 confirm formation of the N phase between the crystalline and the isotropic states. The SAXS peak has *d*-spacing of ~25.0 Å, almost the same as the length of WJ-18 molecule. The wide angle scattering in the vertical direction centers around 4.7 Å, suggesting a slightly larger side-way distance in the N phase. On heating the crystalline sample melts at ~180 °C, whereas on cooling it crystallizes at 130 °C at the rate of 2 °C/min. (Figures. S45 and S46, ESI)

**3.4. Electrical study and discussion of the hysteresis in *P* vs. *E***

The main objective of the molecular design has been to enhance the dipole moment and to increase the dielectric permittivity to understanding the structure-property relationships and its correlation to the ferroelectric nematic phase in related materials such as DIO. We examined ferroelectric behavior or its absence in WJ-16 and WJ-18 through studies of the hysteresis in the polarization *P* vs. *E* plots. The usual method of measuring the polarization as a function of *E* is by Sawyer-Tower circuit [38,39]. In this circuit, the sample is connected in series with a capacitor of a known value thus forming a capacitor divider electrical circuit. To obtain whether the hysteresis in *P* vs. *E* is present, the output of capacitor divider circuit is connected to the Y-input of the oscilloscope whereas the applied triangular voltage is connected to its X-input in the X-Y mode of the oscilloscope. Figures 4A and 4B show electrical responses of the sample to applied triangular wave voltage signal of 100 $V_{pp}$, (subscript 'pp' to V stands for the peak-to-peak voltage) using (A,C) the capacitive and (B,D) the resistive circuits (the latter circuit is shown in the inset of Fig. 4B).

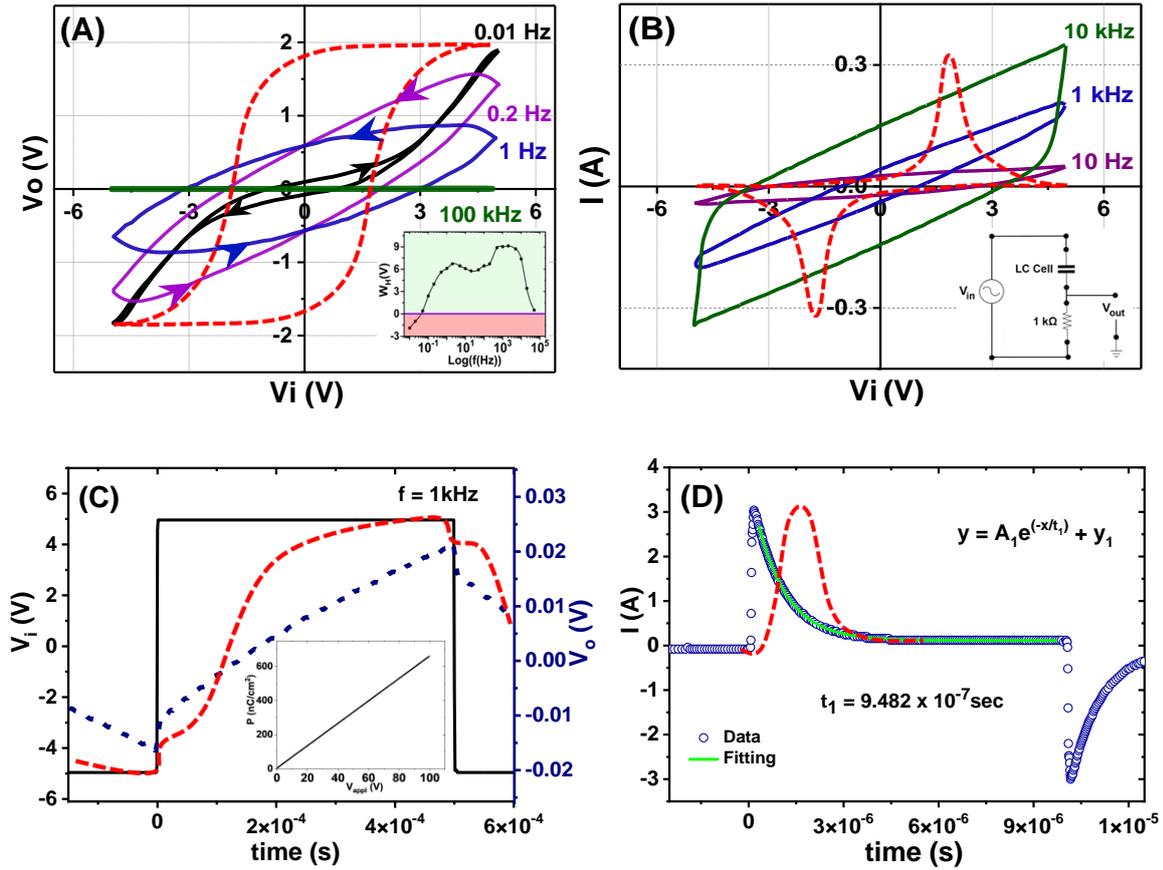

**Figure 4.** The capacitive (**A**) and the resistive (**B**) response of **WJ-16** filled 9 μm uncoated cell at 150 °C for different frequencies, 100 $V_{pp}$ triangular voltage is applied (in the X-Y mode) of the oscilloscope. The inset in Figure 4A shows variation of the measured hysteresis width ($W_H$) as a function of frequency of the applied signal; (**C**) and (**D**) correspond to recorded voltage across the capacitor and the current through the resistance on the application of 100 $V_{pp}$ square voltage, connected to the oscilloscope in Y-t mode. Here, red dashed lines correspond to anticipated ferroelectric hysteresis response, whereas the blue dash lines in C correspond to the actual voltage response and the blue data points in D represent the observed current response. Green line is a fit to the equation given in the inset of Fig. 4D.

The shape of the response curves can be considered "pseudo-hysteresis", i.e., the polarization lags behind the electric field since the negative and the positive slopes do not match with each other. Furthermore, the saturation of *P* with *E* is not exhibited. The proof of the real classical hysteresis behavior is that a saturation in the polarization *P* vs *E* is observed as a dashed curve in Figure 4A. In our case, the shape of the experimental response curve is far from the normal hysteresis curve (observed for DIO [7]). The shapes of such pseudo-hysteresis curves are strongly dependent on the frequency of the applied field, and these shapes can easily be explained by the dynamic lag of the electric field [40]. It is to be noted that the problem of finding 'static hysteresis loop' experimentally is a challenging task and it was studied in detail in Ref. [41]. Firstly, the hysteresis loop width (the double coercive voltage) shows dynamic increase with frequency due to the dynamic delay in switching, [40] hence the frequency for a given applied field must be sufficiently low. Therefore, even with the absence in the saturated polarization, the dynamic curves can easily be denoted as the "pseudo-hysteresis" with hysteresis-like response. Secondly, the hysteresis width (or the double coercive force) continues to vary evenly in the low and ultra-low frequency ranges: from 1 Hz down to $10^{-4}$ [41] due to the screening effect from the charges due to impurities that accumulate on the interface between the alignment layers and the dielectric medium. This effect explains the inverse (negative) hysteresis loop observed at very low frequencies of 0.01 and 0.2 Hz (Figure 4A). Hence no real ferroelectric hysteresis is observed in our samples. This is additionally confirmed by measuring the response using the resistive load (Figure 4B). Ferroelectric switching must lead to two distinctive peaks [42] as shown by dashed lines in Figure 4B. We do not observe these peaks for WJ-16 and WJ-18, clearly proving that the switching behavior is not ferroelectric. For ferroelectric nematic DIO, two current peaks are clearly visible in the $N_F$ phase [21].

An alternative method of measuring the spontaneous polarization, using the circuit with the capacitive [43] and the resistive [9] loads, is to monitor the response across the load to the applied square-wave voltage in time domain (i.e., using the Y-t mode of the oscilloscope). This allows for a simultaneous measurement of both the switching time and the rotational viscosity. Figures 4C and 4D show the observed voltage and the current responses using the capacitive (2 μF) and the resistive (1 kΩ) loads. Instead of the waveform characteristic obtained for ferroelectricity (shown by grey-dashed lines) in terms of the smooth voltage peaks arising from an integration of the re-polarization current [43], we observe classical paraelectric waveform. Similarly, for the resistive circuit, we expect the output to be Gaussian as reported by Chen *et. al*. [9]. However, we observe only the exponential decay of the current but without elongation of the pulse, a typical response from a conventional nematic LC.

Another way to check whether it is ferroelectric or not is a P-E dependence. The inset of Fig.4 shows the dependence of electrical displacement in 4 μm planarWJ-16 cell on applied sine-wave voltage. One may see that that this dependence is perfectly linear in both ways: on increasing and decreasing the applied amplitude up to 100 V (25 V/μm) at which the induced polarization is ~0.7 μC/cm$^2$. We can estimate how far it is from saturation. Assuming a typical (for such high dipole moment) value of spontaneous polarization ~5 μC/cm$^2$ [11,12] the expected saturation will occur at 500 V or higher. Therefore for practical voltage the P-E response is paraelectric-like.

The second harmonic generation (SHG) experiments are performed on these samples. Unlike ferroelectric nematic phases [7] the SHG signal was not observed confirming the absence of ferroelectricity.

Summarizing, neither hysteresis in P vs. E nor any other ferroelectric property was observed in the entire liquid crystalline temperature range of both WJ-16 and WJ-18.

### 3.5. Dielectric spectroscopy

Despite our expectations, both materials do not show any presence of ferroelectricity in in POM, P-E and SHG studies however they show CP. To study the physical origin of this phenomenon we utilized the dielectric spectroscopy, one of the most sensitive techniques for study ferroelectric and other polar materials/phases. This was successfully employed for characterization of ferro-[44,45,46,47] / antiferro- [48,49] and ferri-electric [50,51] liquidcrystalline phases. Dielectric spectroscopy of WJ-16 and WJ-18 are investigated using Novocontrol high precision Impedance Alpha analyzer initially employing commercial cells of different thicknesses with both homeotropic and planar alignment. The dielectric spectra were taken in the frequency range of 0.1 Hz to 10 MHz with 0.1 V measuring AC voltage. The complex permittivity data were fitted to the Havriliak-Negami, eqn. (1) below, where the DC conductivity term is also included [52]:

$$\varepsilon^* = -\frac{i\sigma}{\varepsilon_0 \omega} + \varepsilon_\infty + \sum_{j=0}^{n} \frac{\Delta\varepsilon_j}{\left[1+(i\omega\tau_j)^{\alpha_j}\right]^{\beta_j}} \qquad (1).$$

Here $\varepsilon^*$ is the complex permittivity and $\varepsilon_\infty$ is the high frequency permittivity. The latter includes electronic and atomic polarizabilities of the material. $\omega$ is the angular frequency of the probe field, $\varepsilon_o$ is the permittivity of free space, $\sigma$ is the DC conductivity, $\tau_j$ is the relaxation time, $\Delta\varepsilon_j$ is the dielectric strength of the $j^{th}$ relaxation process, $\alpha_j$ and $\beta_j$ are the corresponding symmetric and the asymmetric broadening parameters of the distribution of relaxation times.

Initial measurements were made in commercial cells with planar alignment. Figure 5D (Inset) shows temperature dependencies of the total dielectric permittivity (i.e. real part of complex permittivity, $\varepsilon'$) of 4 μm and 9 μm planar cells. WJ-16 shows significantly large values of the dielectric strength (~200 and ~450) which could possibly be associated with ferroelectricity. However, it should be noted that the dielectric spectra of WJ-16 are different from the ferroelectric DIO material [53]. The dielectric permittivity WJ-16 is temperature independent of frequency in the entire temperature range, while DIO shows temperature dependent soft-mode like behavior in non-ferroelectric phases (N, SmZ$_A$) and then temperature independent permittivity in ferroelectric nematic phase, $N_F$. Independence of permittivity observed in Fig. 5D on temperature is rather an unusual behavior and it needs further attention. Recently, Boulder group (N. Clark et al.) have found the effect of the insulating alignment layers on the apparent/measured values of the dielectric permittivity [15]. In this case the apparent capacitance $C_{app}$ of LC cell can be expressed as two capacitances in series: capacitance of the LC layer $C_{LC}$ and capacitance of $C_I$ of insulating alignment layer. Hence the apparent $C_{app} = \frac{C_{LC} \cdot C_I}{C_{LC} + C_I}$. There can be two possible opposite cases: the ordinary case where $C_{LC} \ll C_I$, and the extraordinary case, where $C_{LC} \gg C_I$. In the ordinary one, i.e., in materials with the low/moderate dielectric permittivity, capacitance of the LC cell $C_{LC} \ll C_I$ and the apparent capacitance is the capacitance of LC cell, $C_{app} = C_{LC}$. Therefore, in the ordinary case, this gives real value of capacitance and permittivity. However, in materials with very high dielectric permittivity ($\varepsilon > 10000$), such as the ferroelectric $N_F$ phase and bent-core LCs [54,55] the capacitance of LC cell can easily exceed capacitance of the alignment layer, $C_{LC} \gg C_I$. In such a case, the apparent capacitance is limited by the capacitance of the insulating layers, $C_I$. We term it as "Clark's limit", $C_{app} = C_I$. In our case called the extraordinary, apparent $C(T)_{LC} \gg C_I$, the LC capacitance (and dielectric permittivity) would be limited by Clark's limit $C_I$ which is a constant over a reasonable range of temperatures.

Another important feature of the extraordinary case is that the apparent capacitance will be linearly dependent on the cell thickness [15]. Both features of the extraordinary case are supported by the results given in Fig. 5D. We can estimate capacitance of the insulating alignment layers of our cells $C_I = \frac{\varepsilon_I \varepsilon_0 A}{2d_I}$, and assuming the layer thickness $d_I = 200$ nm, and its permittivity as $\varepsilon_I \sim 5$, we obtain $C_I \sim 5.5$ nF. This capacitance limits the apparent value of the dielectric permittivity for 4 μm cell to 100 which is in good agreement with the experimental value of 180, considering that the assumed values of thickness and the permittivity of alignment layer may be different from the assumed values. To avoid this limitation and to get the actual values of the dielectric permittivity, one needs to increase capacitance of the alignment layers $C_I$ in approaching the ordinary case $C_I \gg C_{LC}$. The only way to do this is to reduce their thickness or simply use uncoated metal/ITO electrodes as has been used in some works [11,13,14,16,28, 56] and this allows to measure permittivies up to 50,000.

Following this idea, we performed dielectric measurements of both WJ-16 and WJ-18 in a handmade 4 μm cells with bare ITO. Figure 5 (A) and 5(B) show three-dimensional (3D) plot of the real and imaginary parts of dielectric permittivity of 4 μm WJ-16 uncoated cell. Figure (C) shows the dielectric loss spectra for uncoated 4 m WJ-16 cell taken at 130, 180 and 200 °C with the corresponding fitting at 130 °C and (D) Temperature dependence of the total dielectric permittivity of uncoated 4 WJ-16 and WJ-18.

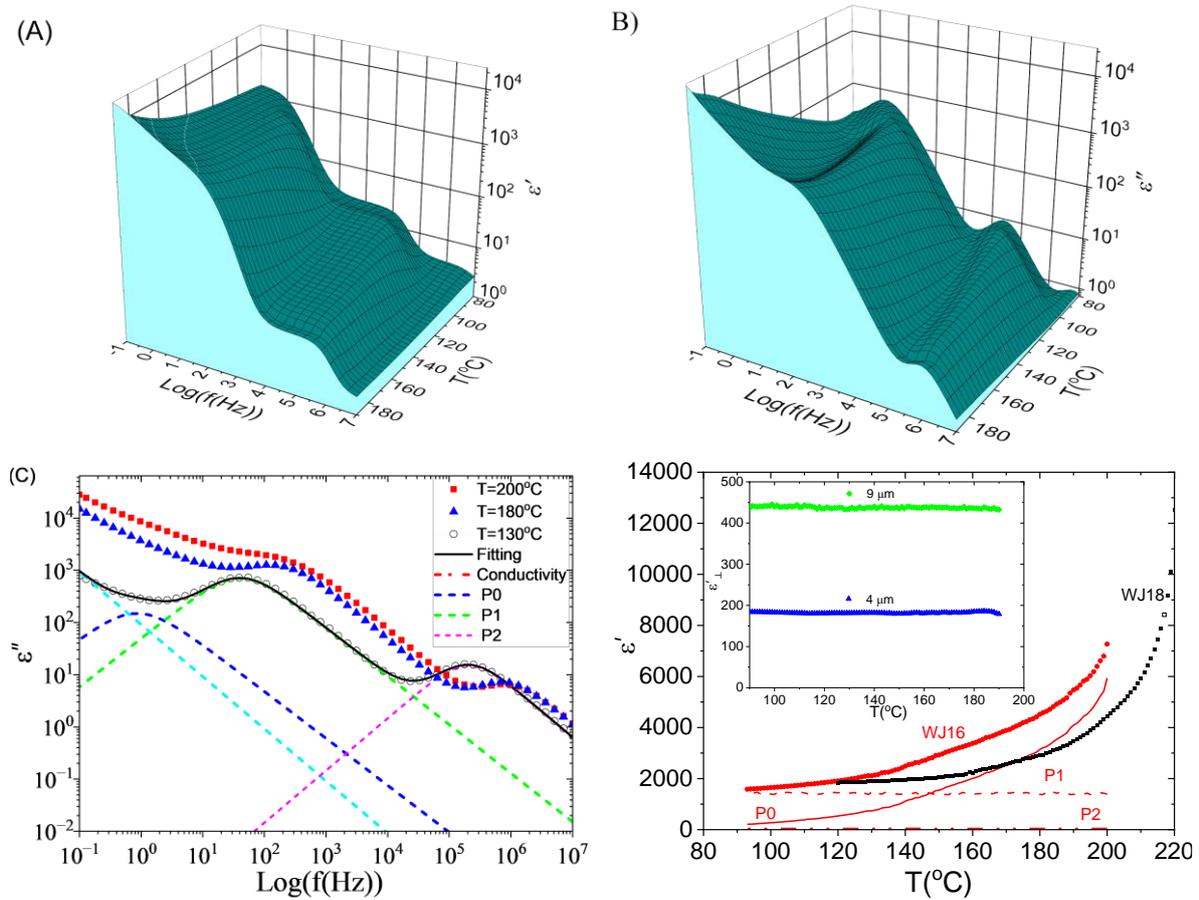

**Figure** 5 **(A) and (B)** show three-dimensional (3D) plot of the real and imaginary parts of dielectric permittivity of WJ-16 4 μm cell with bare ITO electrodes; **(C)** shows the frequency dependence of the imaginary part of permittivity at a temperature of 111 °C; circle symbols, ○, denote the experimental data, the solid black line is a fit to Eq. 1, while the blue, green and magenta dashed lines denote P0 (the electrode polarization process), relaxation processes P1 and P2, while the red dashed straight line with a slope of -1 represents the DC conductivity; **(D)** Temperature dependence dielectric permittivity for two 4 μm uncoated cells for WJ-18 (■) and WJ-16 (●) and dielectric strengths of three individual relaxation processes P0 (solid red line), P1 (dashed red) and P2 (dot-dash). **Inset**: Temperature dependence dielectric permittivity of two WJ-16 planar 4 μm (▲) and 9 μm (◊) cells

A comparison of the temperature dependencies of the total dielectric permittivity of 4 μm planar cell (Fig. 5D) and uncoated 4 μm cells (Fig. 5D) of WJ-16 shows that the dielectric strength of uncoated cell is about one order higher than of the planar aligned cell and becomes temperature dependent. These two features reflect that the Clark's limit is avoided in bare electrodes cell and the measured values should reflect the real value of the dielectric permittivity. One might notice that in uncoated cells the dielectric permittivity is growing with temperature and reach maximal values in the isotropic phase. To understand this unusual feature, we concentrated on the individual relaxation processes which contribute to the total dielectric permittivity.

Figure 5C is an example of the fitting of the dielectric loss spectra to three relaxation processes at 130 °C. The are totally three relaxation processes named as P0-P2 with increase of the relaxation frequency. During fitting the spectra to Eqn. (1) both stretching parameters (α, β) were initially not fixed for all processes and we obtain α=1 and β=1 for both P1 and P2. The stretching parameters values of unity imply that the relaxation these processes are pure Debye. Normally in the conventional calamitic liquid crystals, there exist two individual relaxation processes, one around the short axis often called the 'flip flop mode' (observed at lower frequencies) and the second around the long molecular axis (exhibited at higher frequencies). Since the dielectric strength of P1 is found to be much higher in homeotropic (not discussed in the present manuscript) than in planar-aligned cell, the process P1 can be assigned to flip-flop mode and P2 correspond to the rotations around the long molecular axis, as reported in the literature [20,53,57]. The stretching parameter, $α_0$, for P0 lies in the range 0.65 - 0.9 depending on the temperature. Here P0 is assigned to the parasitic ionic relaxation dynamics, caused by the separation of positive and negative charges, their motions in opposite directions end by their accumulation on the two electrodes. Process P0 in the dielectric loss spectra is broadened and hence α < 1. In other words, P0 is assigned to the space charge/interfacial polarization produced by the mobility of ions and finally their accumulation on to the electrodes.

Now we are ready to explain the huge dielectric permittivity in the isotropic phase. Fig. 5(D) shows the temperature dependence of the total dielectric permittivity 4 μm uncoated cells for WJ-16 and WJ-18 and the dielectric strength of processes P0-P1 for WJ-16 which contribute most to the total dielectric permittivity. The parasitic ionic process P0 is strongly temperature dependent, while the P1 is almost independent. In the low temperature range Δε1 >> Δε0 and the total dielectric permittivity is also almost temperature independent. However, the dielectric strength of P0, Δε0 grows quickly with temperature and at ~152 °C it exceeds the dielectric strength Δε1 and contributes most to the total permittivity. Therefore, the huge dielectric permittivity in high temperature range, including the isotropic phase is obviously due to parasitic ionic process. Similar observation and explanation of the CP in isotropic phase was given by [58,59]

Now ignoring the parasitic ionic process P0 let us examine the relaxation process P1 in greater detail. Most striking feature of this process is a CP (~1500) which is independent of temperature and phase but linearly depends on the cell thickness, i.e dielectric strengths in 4, 9, 22 μm cell are 1500, 3000, 7000 correspondingly.  The magnitude of dielectric strength of P1 is two-three orders of magnitude higher than in paraelectrics  and about one order smaller than the Goldstone mode in ferroelectric nematics [7,8,27,28,32,57] and some bent-core LCs [54,55,[60]].  In these ferroelectric materials the dielectric strength strongly depends on bias voltage which suppresses Goldstone mode [13,15,55,].   However our samples unlike the ferroelectric materials show paraelectric-like, hysteresis free linear P-E response and this property together with CP can be assigned to SPE which appears in the temperature range between paraelectric and ferroelectric phases [[61]]  As a result of this and in contrast to ferroelectric nematics the dielectric spectra of our materials do not depend on bias voltage due to the absence of spontaneous polarization.

High dielectric permittivity very likely reflects extremely strong collective molecular dipolar contribution.  In solid SPE materials this is a contribution from well-aligned molecular clusters which are also exist in bent-core LCs [54,55].  In our nematic materials such cluster were not observed and the CP can be explained by response of correlated conglomerates of molecules.  The possible physical mechanisms will be discussed later on.

Since the results of the permittivity measurements by dielectric spectroscopy are currently hotly debated [14,15,17], to validate the obtained permittivity values, we use direct capacitance measurements to support observation of the colossal permittivity CP.

### 3.6. Direct capacitance measurements.

The colossal dielectric permittivity (CP) obtained by dielectric spectroscopy of uncoated cells can be confirmed by the direct capacitance measurements of the LC sample.  By definition the dielectric permittivity, ε' is a ratio of capacitances of the filled and empty cell, so the direct capacitance measurements will give most reliable value for capacitance and permittivity irrespective of any physical model.   We used a capacitor divider circuit, shown in the inset of Figure 7, the capacitive load, $C_0 = 2$ μF.  The capacitance of LC sample is measured as $C_{LC} = C_0 \frac{V_0}{V_{in} - V_0}$, the real part of the permittivity is deduced from $\varepsilon'_{LC} = \frac{d \cdot C_{LC}}{\varepsilon_0 A}$. Figures 7A and 7B show frequency dependencies of the dielectric permittivity for different temperatures (SmA, N and the Iso phases) using 4 μm uncoated cells of WJ-16 and WJ-18, respectively.

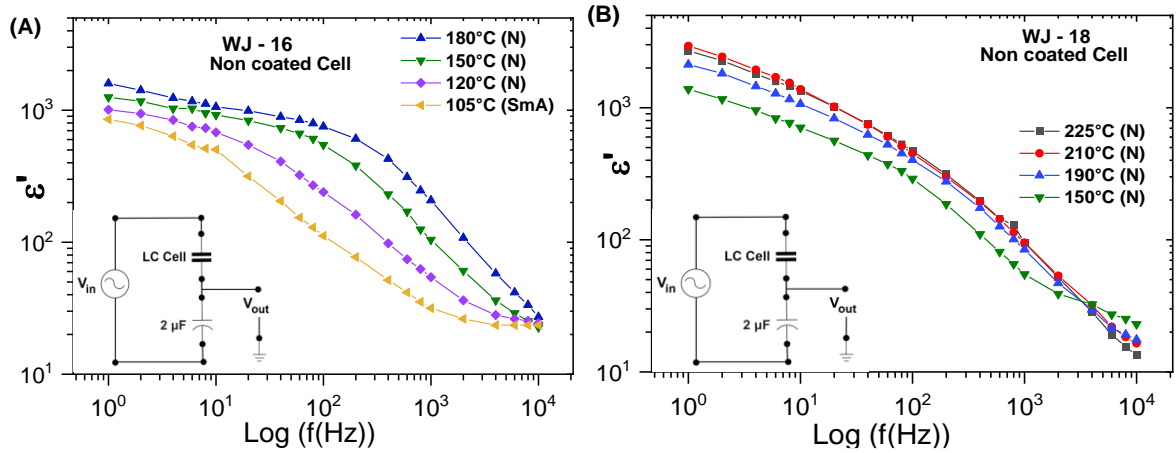

**Figure 6.** The direct measurement of dielectric permittivity (ε′) as a function of frequency for different temperatures using the capacitance voltage divider circuit using uncoated 4 μm cells filled with (A) WJ-16 and (B) WJ-18. The inset in figure 6B shows a simple capacitance divider circuit used in the measurement of ε′, where *Vin* is applies sin-wave voltage and *Vout* in the output voltage taken from the output 2 μF capacitor.

It is interesting to note that frequency dependence of the dielectric permittivities calculated from the direct capacitance measurements in uncoated cells of both samples (Figure 7A,B) match with the values measured by dielectric spectroscopy (Figure 5). The result demonstrate CP in both N and SmA phases.

A direct measurement of the dielectric permittivity provides a clear and direct evidence of the observation of the colossal dielectric permittivity in the non-ferroelectric compounds, WJ-16 and WJ-18. Hence, this behavior can be related to superparaelectricity (SPE) [**Error! Bookmark not defined.**,5] in soft matter systems. Superparaelectricity reflects an incomplete ferroelectric long-range ordering. Ferroelectric state is an assembly of long-range-ordered polar domains having the spontaneous polarization (*Ps*) and it exhibits hysteresis in the polarization as function of the external electric field, while the superparaelectric state consists of the short-range polar order clusters. For the latter, the macroscopic polarization <Ps> ~ 0 and hysteresis free P–E curve is observed here. Figure 7 is a schematic of the proposed dipolar structures in WJ-16 for the three phases. Both states, superparaelectric and the ferroelectric, show large dielectric permittivity.

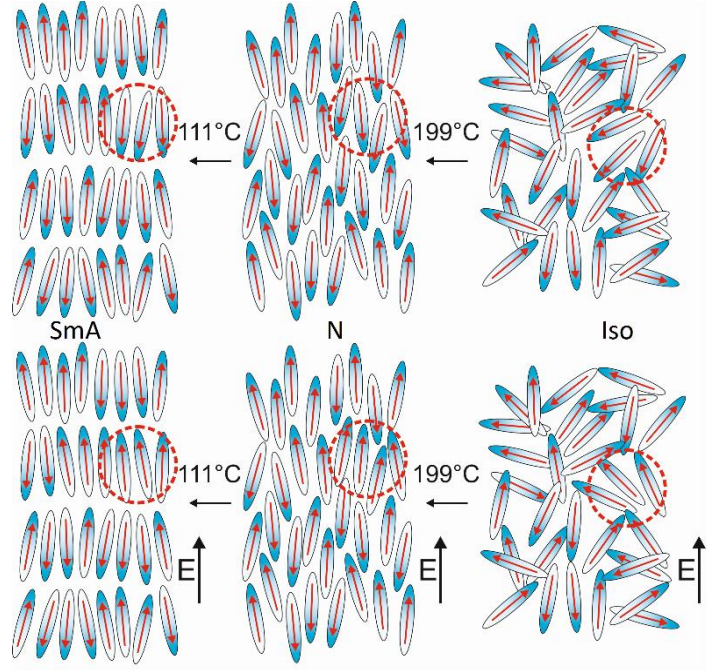

**Figure 7.** The proposed schematics of dipolar structures and the molecular assembly in the phase sequence of WJ-16.

However, our studied compounds W-16 and Wj-18 exhibit colossal permittivity even in the isotropic phase with the absence of hysteresis in all phases. Therefore, the phases of these compounds can be assigned superparaelectrics rather than ferroelectrics. The higher values of dielectric permittivity can be explained using the classical approach suggested by Kirkwood and Frohlich, [62] where the static dielectric permittivity $\epsilon_S$ in paraelectric isotropic liquids is given by:

$$\epsilon_S - n^2 = \frac{3\,\epsilon_S}{2\epsilon_S + n^2} \frac{N}{3\,kT\epsilon_0}\, \mu\mu^* \qquad (2)$$

$n$ is the refractive index at a visible wavelength, $kT$ is the thermal energy, $\mu$ is the dipole moment of the molecule, $\epsilon_0$ is the permittivity of free space. $\mu^*$ is the average moment of the dipoles in a macroscopic spherical region surrounding the dipole on the assumption that one of the dipoles is kept in a fixed direction. $N$ is the number density of molecules. $M/d$ is the molar volume, $N_A$ is the Avogadro number; $N \cdot \frac{M}{d} = N_A$ and $\mu\mu^* = \mu^2 (1 + z\,\overline{\cos\gamma})$, where $\overline{\cos\gamma}$ is the average of the cosine angle between the neighboring molecules, $z$ is the average number of the interacting dipoles, also called the coordination number. Assuming $\overline{\cos\gamma}$ equal to unity, the dielectric permittivity is proportional to the square of the dipole moment and to the coordination number, $z$ plus one. A ferroelectric copolymer with the permittivity of 130 has a coordination number of 30 [63], while the coordination number of 1000 was reported in a bent core compound [55**Error! Bookmark not defined.**].

In solid SPE the dielectric permittivity is temperature dependent because it depends on the size of ferroelectric domains/clusters but not on the size of the sample. Contrary to this the dielectric strength n ferroelectric LCs is proportional to the cell and this observed in different phases such as surface-stabilized SmC* [47], bent-core LCs [55] and ferroelectric nematic $N_F$ [15,16]. This implies that liquid crystals have a long-range directional order and the correlation length is limited by the cell thickness. Therefore the molecular dynamics can be modeled with boundary condition at the electrodes which gives a linear dependence of dielectric strength on the cell thickness [15,47,55,58]. Preliminary study of our materials shows that the dielectric strength of SPE relaxation process P1 is proportional to the thickness of the cell (*d*) and the high permittivity in our systems can be attributed to the collective fluctuations of a very large number of molecules having the correlation length ~ *d* similar to the other ferroelectric LCs mentioned above. However there is big difference in dielectric spectroscopy of FLCs and our materials. The dielectric strength of ferroelectric materials is strongly depends on bias voltage suppressing the ferroelectric Goldstone response [13,15,55,58]. Contrary to this, the dielectric strength of our materials is practically independent of bias voltage up to 35 V available in Novocontrol Alpha Analyzer. Although the dielectric permittivity in $N_F$ and in our sample linearly depends on the cell thickness they have different physical origin, the phason mode in $N_F$ and amplitude mode in SPE, which needs further study.

## 4. Conclusion

In this study, we have designed, synthesized, and investigated two novel compounds WJ-16 and WJ-18, as modifications of the prototype DIO. The dipole moments determined using DFT are 10.4 and 14.8 D, respectively. Our original expectation was development of new ferroelectric nematic materials. Surprisingly, both materials show linear, non-ferroelectric, hysteresis-free P - E dependence. Recently R. Mandle [33] showed that the high magnitude of the dipole moment does not guarantee formation of ferroelectric nematic phases. Madhusudana [30] theoretically showed the importance of charge distribution along the molecule for formation ferroelectric state. Finally Li *et al* [32] systematically analyzed more than 100 ferrogenic compounds and defined the Pearson's coefficient (or impact) of different parameters responsible for ferroelectric nematic phase. The most important are the dipole moment (0.26), molecular length (0.19) and the dipole angle (0.16). The dipole moment itself being most important however cannot guarantee the ferroelectric nematic phase. Both samples being non-ferroelectric show CP measured using a dielectric spectrometer as well as by direct capacitance measurements of the LC cell with uncoated electrode. Hence, the observed phenomenon is best described as superparaelectric rather than ferroelectric. This is the first direct demonstration of superparaelectricity in liquid crystalline organic materials. The results given in this manuscript highlight the anisotropic properties of the large dipole moment soft matter organic molecular system. The synergistic electrical properties are subtle, interesting, and potentially useful.

## Acknowledgment

The work in Dublin was funded by the US-Ireland, SFI 21/US/3788.. WJ thanks the CSC, China for a PhD scholarship. YT thanks Faculty of Engineering, University of Sheffield for his PhD studentship. MS thanks Prof. Manjuladevi V, BITS Pilani, India for allowing him to work as visiting student at Trinity College Dublin. The work in Sheffield was funded by EPSRC (EP-T003294). GHM thanks Diamond Light Source, UK for beamtime allocations SM31552 and SM33389. XZ thanks Dr. Olga Shebanova and Prof. Nick Terill at station I22, Diamond Light Source for help with SAXS/WAXS experiments.